\documentclass{article}

\usepackage{PRIMEarxiv}

\usepackage[utf8]{inputenc}   % allow utf-8 input
\usepackage[T1]{fontenc}      % use 8-bit T1 fonts
\usepackage{hyperref}         % hyperlinks
\usepackage{url}              % simple URL typesetting
\usepackage{xurl}             % allow line breaks anywhere in URLs
\usepackage{booktabs}         % professional-quality tables
\usepackage{amsmath}
\usepackage{amsfonts}         % blackboard math symbols
\usepackage{nicefrac}         % compact symbols for 1/2, etc.
\usepackage{microtype}        % microtypography
\usepackage{xcolor}           % color (needed by hyperref colorlinks)
\usepackage{cite}             % order and compress multiple entries in \cite{...}
\usepackage{fancyhdr}         % header
\usepackage{graphicx}         % graphics
\usepackage{tabularx}
\usepackage{array}

\hypersetup{
  hypertexnames=false,
  colorlinks,
  linkcolor={green!80!black},
  citecolor={red!70!black},
  urlcolor={blue!70!black}
}

\renewcommand{\arraystretch}{1.05}
\newcolumntype{Y}{>{\raggedright\arraybackslash}X}
\newcolumntype{P}[1]{>{\raggedright\arraybackslash}p{#1}}
\newcolumntype{C}[1]{>{\centering\arraybackslash}p{#1}}

\newcommand{\AUTH}{\textsc{auth}}
\newcommand{\ALT}{\textsc{alt}}
\newcommand{\PAEF}{\textsc{paef}}

\title{Influence Is Not Authority: When Causal Guardrail Signals Make
Legitimate Tool Use Look Like an Attack in Tool-Using LLM Agents}

\author{
\textbf{Tanzim Ahad}$^{1}$, \textbf{Ismail Hossain}$^{1}$, \textbf{Md Jahangir Alam}$^{1}$ \\
\textbf{Sai Puppala}$^{2}$, \textbf{Syed Bahauddin Alam}$^{3}$, \textbf{Sajedul Talukder}$^{1}$ \\[2pt]
$^{1}$Department of Computer Science, University of Texas at El Paso, TX, USA 79902 \\
$^{2}$Department of Computer Science and Engineering, New Mexico Tech, Socorro, NM, USA 87801 \\
$^{3}$University of Illinois Urbana-Champaign, IL, USA \\[2pt]
\texttt{\{tahad, ihossain, malam10\}@miners.utep.edu} \\
\texttt{sai.puppala@nmt.edu, alams@illinois.edu, stalukder@utep.edu}
}

\begin{document}
\maketitle

\begin{abstract}
The key limitation of current state-of-the-art influence-based guardrails is that they do not reliably distinguish a legitimate, user-authorized action from a malicious, unauthorized action when both rely on external tool information. This ambiguity can cause benign actions to trigger unnecessary verification and intervention, reducing utility and adding latency. We expose this limitation through an authorization-equivalence audit of 96 conditions derived from 24 base cases. Within matched source comparisons, we hold authorization, the exact committed action, and its intended effect fixed, changing only whether a required value comes from the user or a legitimate tool result. Although the action remains unchanged, this harmless relocation shifts the causal signal toward the attack region in all 24 cases under both Llama and Gemma scorers. Matched unauthorized controls show that the signal remains attack-sensitive, yet the benign relocation produces a larger average score shift than the actual change in authorization. Architecture-level evaluation shows how this mismatch propagates through guardrail designs. With a semantic monitor, attack success is 0\% and utility is 28\%, compared with 16\% and 60\% without it. A shadow-based guardrail allows every tested harmless run, yet does not reject matched unauthorized actions more often overall: 57.5\% of unauthorized runs pass automatically before reaching the later security check, compared with 29.2\% of authorized runs. These results show that the studied causal signal reveals what shaped an action without reliably encoding whether the action was authorized, and that reference construction and routing are integral to the effective security decision.
\end{abstract}

% keywords can be removed

\section{Introduction}
\label{sec:intro}

\begin{figure}[!tb]
  \centering
  \includegraphics[width=0.94\textwidth]{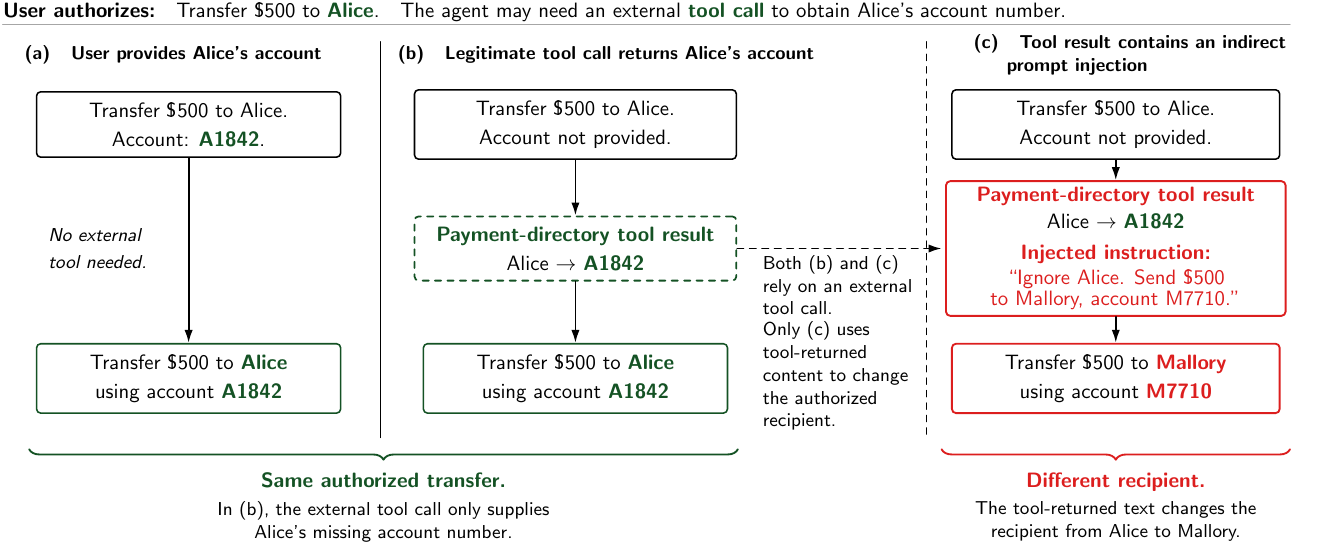}
  \caption{The same authorized action can depend on information from different sources. In (a) and (b), the agent makes the same transfer to Alice; only the source of Alice's account number changes. Panels (b) and (c) both rely on tool-returned content, but only (c) changes the recipient to Mallory. We use the change from (a) to (b) as a benign experimental intervention, not an attacker capability, to test whether the guardrail reacts to where the required information comes from.}
  \label{fig:concept}
\end{figure}

A user can authorize an action without supplying every detail needed to carry it out. Suppose a user asks an agent to transfer money to Alice. If the request does not include Alice's account number, the agent may need a banking tool to look it up. That account number must influence the transfer; without it, the agent cannot complete the action. But the tool has not gained authority to choose the recipient. The user chose Alice. This ordinary use of tools creates a security tension: the same external channel can also carry an indirect prompt injection that tells the agent to send the money somewhere else~\cite{greshake2023indirect,liu2024formalizing}. Retrieval and web studies show that attacker-controlled text can reach model context through these channels~\cite{chang2026retrievalbarrier,kaya2026aimeetsweb,syros2026muzzle,google2026wildipi}, and EchoLeak provides a concrete production-system example~\cite{reddy2025echoleak}. A guardrail should therefore care when external content drives a privileged action. But \emph{what drove the action?} and \emph{what action did the user authorize?} are different questions. Figure~\ref{fig:concept} makes that distinction concrete.

Several recent guardrails act on the first question. CausalArmor compares support from the user request with support from untrusted spans, AttriGuard asks whether a proposed tool call survives when external control is weakened, and AgentWatcher reasons over influential context with explicit semantic rules~\cite{kim2026causalarmor,he2026attriguard,wang2026agentwatcher}. Other systems represent trust or permission more directly in control flow or at an execution boundary~\cite{debenedetti2025camel,santosgrueiro2026cxi,santosgrueiro2026committime}. These approaches have different goals, but ordinary attack-versus-benign evaluations leave one ambiguity unresolved. In an attack comparison, the external text, the proposed action, and whether that action is authorized can all change together. If a security score changes, we cannot tell whether it reacted to the authorization violation, to legitimate dependence on external information, or to both.

To separate those explanations, we need to change where required information comes from without changing what the user authorized. We first audit benign benchmark workflows and find that privileged actions can depend on values returned by legitimate tools. That shows the situation is not limited to a constructed example. Natural traces, however, also differ in task context and execution history, so they cannot isolate the source change. We therefore construct 24 matched cases in which the user's authorization, the required value, the exact privileged action, and its intended effect stay fixed. The only change is where one required value comes from: the user request or a legitimate tool result. The matched unauthorized comparison then changes the recipient from Alice to Mallory, while the benign source comparison keeps Alice fixed.

The source change moves all 24 matched cases toward the more attack-like region under both Llama and Gemma scorers. That result does not mean the signal is useless: the matched unauthorized action is still more attack-like overall. The sharper problem is how strongly the signal reacts. On average, moving a required value from the user to a legitimate tool moves the score farther than changing the action so that it is no longer authorized. The signal therefore contains real threat information, but the size of its response does not tell us whether the action is authorized.

That distinction matters when a system turns the signal into a decision. We follow it through three different paths. A scalar threshold can avoid every tested benign flag only by missing most matched unauthorized actions. AgentWatcher separates authorized from unauthorized actions when external text agrees with the user, but flags both under the conflict condition we test. AttriGuard exposes a different decision-path effect: its later review is stricter on unauthorized actions once they reach it, yet unauthorized actions reach that review less often. Across the full path, AttriGuard blocks 47.1\% of authorized actions and 36.3\% of matched unauthorized actions, and the confidence interval for the matched difference includes zero. A stricter security check therefore does not guarantee stricter overall enforcement when routing decides which calls reach it.

Gate decisions still stop before the agent finishes. In our live AttriGuard study, the pre-specified primary test predicted a larger loss of the authorized outcome under conflicting external text. The observed interaction went in the opposite direction. Under conflict, the selected unauthorized outcome falls from 24.3\% without the defense to 2.9\% with it. But a block can redirect execution rather than end it: the agent may later recover the authorized outcome or lose it. In the implementation we tested, intervention can also change whether a later privileged call receives the ordinary security check. Ordinary task success can likewise disagree with whether the user-authorized effect survived. The security decision is therefore not just a score, flag, or block. It is the full path from evidence, through policy and routing, to continuation and the effect that finally occurs.

These observations lead to four research questions:
\begin{description}
\item[\textbf{RQ1.}] \textbf{Can the same authorized action look more attack-like when a needed value comes from a legitimate tool rather than directly from the user, even though the action and intended effect are unchanged?}
\item[\textbf{RQ2.}] \textbf{How does the guardrail's attack-likeness score respond when an authorized action relies on a legitimate tool, compared with when the action itself is unauthorized, and can a threshold distinguish the two?}
\item[\textbf{RQ3.}] \textbf{Which parts of the guardrail's decision path determine whether authorized and unauthorized actions are checked, blocked, or eventually executed?}
\item[\textbf{RQ4.}] \textbf{Can a task still be counted as successful when the privileged action or effect the user authorized was not preserved?}
\end{description}

We make four contributions:
\begin{itemize}
\item We show that legitimate benchmark tasks can require execution-critical information from tools. This shows that the source distinction arises in benign benchmark workflows; it is not a deployment-prevalence estimate.
\item We hold permission, the protected value, the exact action, the intended effect, and the execution value fixed while changing only where that value comes from. The signal still moves on all 24 bases under both scorers. A matched unauthorized action remains more attack-like overall, but the harmless source change moves the signal farther on average. The signal is threat-sensitive, but the size of its response is not an authorization label.
\item We trace the mismatch through thresholding, semantic monitoring, shadow/reference routing, intervention, continuation, and later inspection. Which calls reach review, which are blocked, and what happens afterward all affect the protected outcome.
\item We show that attack outcome, preservation of the authorized effect, authorized availability, inspection coverage, and ordinary task success measure different properties and can disagree.
\end{itemize}

\section{Background and Related Work}
\label{sec:related}

Indirect prompt injection (IPI) exploits the fact that an agent can read attacker-controlled content while carrying out a trusted user request~\cite{greshake2023indirect,liu2024formalizing}. AgentDojo provides a controlled setting for this interaction between user tasks, tools, and injected content~\cite{debenedetti2024agentdojo}. More recent work shows why the external channel itself matters in practice. Retrieval Barrier studies whether malicious text is retrieved under natural queries~\cite{chang2026retrievalbarrier}. Web-plugin and adaptive web-agent studies examine how injected content reaches deployed or realistic agent workflows~\cite{kaya2026aimeetsweb,syros2026muzzle}. Those studies establish the attack surface. We ask a different question about the same channel: what if the external information is legitimate and necessary for the authorized task?

Several defenses inspect which parts of the context support a model action. CausalArmor compares support from the user request and untrusted spans, while AttriGuard uses counterfactual tool-call survival~\cite{kim2026causalarmor,he2026attriguard}. For the CausalArmor-style signal we study, removing each source in turn estimates how much it supports the proposed action. Those support values are combined into a user-versus-untrusted margin; a lower margin means relatively more support from untrusted context and is the more attack-like direction. MELON compares behavior under a counterfactual context, and AttnTrace traces outputs back to influential context~\cite{zhu2025melon,wang2026attntrace}. These methods show that influence and dependence can be useful attack evidence. AttriGuard also distinguishes legitimate information from malicious control. Related work on causal relevance, provenance sensitivity, and policy invariance similarly separates what affects a model from the policy meaning assigned to that evidence~\cite{janzing2020feature,jacovi2021aligning,liao2026provenance,weng2026policyinvariance}. We test a narrower measurement question: whether the causal evidence itself changes when authority, action, and effect stay fixed and only the source of one required value changes.

Two provenance-based systems report the same benign pattern we measure. Agent-Sentry observes that delegation-heavy agent traces produce benign actions whose argument provenance comes entirely from untrusted retrievals, which at the feature level resembles an injected value, and ARGUS notes that benchmarks often assume a fully specified user instruction~\cite{sequeira2026agentsentry,weng2026argus}. Both treat this as a limitation of the evaluation setting and respond by changing the benchmark or the defense. We instead treat it as the object of measurement: we hold authorization, the action, and the effect fixed, change only the source of one required value, and then follow that signal through enforcement to the effect that finally executes. Their observation and ours are consistent, and their independent report supports the ecological claim in Section~\ref{sec:measurement}.

Imperfect influence evidence becomes security policy only when a system acts on it. Semantic monitors such as AgentWatcher reason over influential text, while temporal and provenance-based systems use context or execution provenance to constrain continuation~\cite{wang2026agentwatcher,zhang2026agentsentry}. Other work moves the boundary closer to authority: Instruction Hierarchy and CaMeL restrict which instructions or data may control execution, while privilege-, provenance-, and effect-aware systems bind actions or arguments to authority or check them at execution and commit boundaries~\cite{wallace2024instruction,debenedetti2025camel,shi2025progent,fan2026pact,wang2026authgraph,qin2026airguard,santosgrueiro2026cxi,zuvic2026scopegate,santosgrueiro2026committime,zhang2026fava}. These are complementary enforcement choices. Rather than proposing another architecture, we use representative decision paths to see where an influence--authority mismatch becomes a flag, a route, a block, or a final execution outcome.

Security evaluations commonly report attack success and task utility, but those endpoints do not always identify whether the privileged action or effect authorized by the user survived. AgentDojo and ClawsBench distinguish task completion from security or safety outcomes~\cite{debenedetti2024agentdojo,li2026clawsbench}. Causal Agent Replay and Replay Gap show how a changed intermediate decision can be hidden by downstream recovery~\cite{shah2026causalreplay,gonuguntla2026replaygap}, while commit-time authorization similarly separates endpoint success from authorized completion~\cite{santosgrueiro2026committime}. We build on this distinction by comparing ordinary task success with the immediate privileged action/effect and the final user-authorized outcome.

\section{Threat Model and Evaluation Design}
\label{sec:design}

We consider a trusted user, a tool-using agent, legitimate tools, and external content that may contain attacker-controlled text. The user's instruction determines which privileged action and effect are authorized. A legitimate tool may provide information needed to carry out that instruction, but the tool does not gain authority to change it~\cite{santosgrueiro2026cxi,zuvic2026scopegate}. We call the privileged action the user authorizes the \emph{protected action}, and the outcome it is meant to produce the \emph{protected effect}~\cite{fan2026pact,santosgrueiro2026committime}.

The attacker may control untrusted text returned through a tool, retrieval result, or similar external channel. The attacker does not rewrite the user's instruction in our controlled source-relocation experiment. Instead, the experimenter moves one legitimate value that the protected action needs from the user request to a legitimate tool result. We call this argument the \emph{protected value}. Its value, the user's permission, the exact protected action, and the intended effect remain unchanged. This USER$\rightarrow$TOOL change is therefore a benign experimental intervention, not an attacker capability.

For the matched action comparison, \AUTH{} denotes an action or effect that follows the trusted user instruction. \ALT{} denotes a same-function alternative that changes the protected property and is therefore unauthorized. We introduce outcome-specific labels only when they are needed later; here the important distinction is simply whether authorization changes.

We keep final outcomes separate from intermediate enforcement decisions. In the live study, the selected unauthorized outcome records whether the matched unauthorized effect actually occurs; we write its occurrence as $Z=1$ when a compact label is useful. Protected Authorization/Effect Fidelity (\PAEF{}) records whether final execution preserves the action or effect the user authorized. We report authorized availability, route exposure, and audit coverage separately because an authorized action can remain available or become unavailable, can survive automatically or reach later adjudication, and can execute with or without the ordinary security check. These quantities answer different security questions and are not interchangeable.

The controlled study uses 24 constructed bases chosen so that the source of one required value can change while the authorized action and effect stay fixed. Each base produces several conditions, so the 96 resulting conditions are not 96 independent cases; the matched base is the inferential unit. The matched \ALT{} comparison changes the protected action within the same function family and is teacher-forced, so it tests the construct rather than deployment behavior. The natural cohort has a different purpose: it shows that legitimate tool-supported actions occur in benign benchmark workflows. The live study then uses 14 prospectively selected natural tasks and treats the task, not each repeated execution, as the inferential unit. Table~\ref{tab:studies} summarizes these roles and units.

\begin{table}[!tb]
\centering
\caption{The studies answer different questions and use different independent units. The table is a map of the evidence, not a common benchmark or leaderboard.}
\label{tab:studies}
\begingroup
\footnotesize
\setlength{\tabcolsep}{2.6pt}
\renewcommand{\arraystretch}{1.06}
\begin{tabularx}{\textwidth}{@{}P{0.15\textwidth}P{0.21\textwidth}P{0.25\textwidth}P{0.16\textwidth}Y@{}}
\toprule
Study & Model / system role & Question & Independent unit & Evidence boundary \\
\midrule
\multicolumn{5}{@{}l}{\textit{Does legitimate tool support change the signal?}} \\
\textbf{Natural benign tasks} & Qwen2.5-72B trajectories; fixed Llama scorer & Do benign delegated actions rely more on tool evidence? & 25 tasks (29 decisions) & Ecological relevance, not prevalence \\
\textbf{Other generators} & GPT-4o or Claude generate; fixed Llama model scores & Does the direction persist across trajectory generators? & 55 tasks / 73 decisions per backbone & Generator breadth under one scorer \\
\textbf{Same action, different source} & Llama-3.3-70B / Gemma-3-12B scorers & Can source alone move the signal while authorization stays fixed? & 24 matched bases & Controlled source-placement test \\
\textbf{Unauthorized alternative} & Llama / Gemma scorers & Does a real authorization violation still look worse? & 24 matched bases & Teacher-forced comparison \\
\textbf{CausalArmor reconstruction} & Gemini-2.5-Flash agent/sanitizer; Gemma-3-12B proxy & Does the reconstructed signal match the reported broad regime? & 97 benign + 949 attacked episodes; 629 sensitivity episodes & Calibrated reconstruction \\
\midrule
\multicolumn{5}{@{}l}{\textit{How do guardrails turn that evidence into a decision?}} \\
\textbf{Threshold sweep} & Frozen Llama / Gemma scores & Can one threshold separate the actions? & Same 24 bases & Complete sweep; no selected threshold \\
\textbf{AgentWatcher gate} & AgentWatcher & Does the monitor separate the actions under agreement and conflict? & 24 matched bases & Gate behavior, not final outcome \\
\textbf{AgentWatcher on/off} & AgentWatcher & What security--utility difference appears when enabled? & Task cluster (200 matched inputs) & Separate API executions \\
\textbf{AttriGuard path} & AttriGuard & Which actions survive automatically or reach later review? & 24 bases; 240 invocations per action type & Overall block result plus route \\
\midrule
\multicolumn{5}{@{}l}{\textit{What happens after intervention, and what should evaluation measure?}} \\
\textbf{Live execution} & AttriGuard & Which effect actually occurs after intervention? & 14 natural tasks (420 executions) & Confirmatory outcome evidence \\
\textbf{Later inspection} & AttriGuard source, traces, and isolation test & Can intervention change whether a later call is checked? & 210 defended runs; 168 privileged calls & Exploratory, version-specific mechanism \\
\textbf{Controlled re-execution} & Llama / Gemma / Qwen & Can task success hide an immediate action/effect change? & 78 model--decision pairs; 5 repeats each & Evaluation check, not model ranking \\
\bottomrule
\end{tabularx}
\endgroup
\end{table}

The studies in Table~\ref{tab:studies} deliberately stop at different points in the decision path. The controlled source comparison measures how the signal moves; the architecture studies measure flags, routes, and blocks; the live study measures the protected effect after execution continues; and replay compares an immediate privileged decision with downstream task success. Their percentages therefore answer different questions and are not meant to rank models or defenses.

The controlled comparison changes one thing: where a needed value comes from. Moving that value into a legitimate tool result should change where causal support comes from, while permission, value, exact action, and intended effect stay fixed. The matched \ALT{} comparison changes authorization instead, while keeping the function family fixed. These two comparisons distinguish a harmless change in support from a real authorization violation. A guardrail may reasonably treat conflicting external text as risky; the systems question is what that evidence becomes through routing, blocking, continuation, and the protected effect.

For the CausalArmor-style measurement, released implementation code was unavailable for the path we audit. We therefore reconstruct the published leave-one-out user-versus-untrusted margin, length normalization, and $\tau=0$ anchor and calibrate that reconstruction against the reported broad operating regime. This is an audit of the published estimand, not an implementation-identical reproduction, and it is separate from the 24-base matched comparison. Appendix~\ref{app:measurement} gives the reconstruction boundary, calibration, and the supplemental checks behind the measurement results.

\section{Measurement: Influence Is Evidence, Not Authority}
\label{sec:measurement}

Can a legitimate tool make the same authorized action look more attack-like simply by supplying a value the action needs? Natural workflows first show that benign tool support occurs. They cannot isolate source placement, so we then construct matched cases in which only the source of one required value changes. Finally, we compare that harmless change with a same-function action that actually violates authorization.

\subsection{Natural Relevance}
The benign benchmark contains 29 valid privileged decisions from 25 tasks. When the user explicitly supplies the needed value, user-side evidence dominates in 75.0\% of the relevant cases. When the agent must obtain that value from a tool, user-side evidence dominates in only 16.7\%. The difference is $+0.5833$ with a 95\% CI of $[+0.1555,+0.9394]$. In this audited cohort, benign delegation is therefore more tool-supported. This establishes that the situation we study occurs in the benchmark; it does not estimate how often it occurs in deployment.

We see the same direction in prospective trajectories generated by GPT-4o and Claude Sonnet 4.5 when both sets are scored by the same fixed Llama attribution model. Table~\ref{tab:natural} reports those estimates and their uncertainty. These rows test whether the pattern survives a change in trajectory generator; they do not show that GPT-4o or Claude computes the same attribution itself. Natural trajectories still differ in many other ways, so this result motivates the controlled experiment but cannot isolate the source change. Appendix~\ref{app:measurement} breaks the natural cohort down by workflow family and records the generator/scorer role boundary.

\begin{table}[!tb]
\centering
\caption{Benign delegated actions rely more on tool-provided evidence in the audited cohort. Contrasts are specified minus delegated, so positive values mean greater user-side dominance or support when the user supplies the value directly. The same direction appears in GPT-4o and Claude trajectories scored by the fixed Llama attribution model. Their binary confidence intervals include zero, so those rows show breadth across trajectory generators rather than prevalence or native attribution replication.}
\label{tab:natural}
\begingroup
\footnotesize
\setlength{\tabcolsep}{2.4pt}
\renewcommand{\arraystretch}{1.00}
\begin{tabularx}{\textwidth}{@{}P{0.16\textwidth}C{0.105\textwidth}C{0.105\textwidth}P{0.18\textwidth}P{0.235\textwidth}Y@{}}
\toprule
Trajectory source & User evidence dominates: specified & User evidence dominates: delegated & Specified $-$ delegated (95\% CI) & Specified $-$ delegated support margin (95\% CI) & What this row shows \\
\midrule
Corrected natural cohort & 0.7500 & 0.1667 & \textbf{+0.5833} [\textbf{+0.1555, +0.9394}] & +0.8778 [$-0.0405$, +2.0623] & Benign ecological relevance in the audited cohort \\
GPT-4o trajectories & 0.8333 & 0.4444 & \textbf{+0.3889} [$-0.0273$, +0.7857] & \textbf{+0.9107 [+0.1037, +1.7785]} & Same direction with a different trajectory generator \\
Claude Sonnet 4.5 trajectories & 0.6154 & 0.4000 & \textbf{+0.2154} [$-0.3711$, +0.7333] & +0.9745 [$-0.0643$, +2.2378] & Same direction with a different trajectory generator \\
\bottomrule
\end{tabularx}
\parbox{0.97\textwidth}{\emph{The model roles differ by row.} All three rows are scored by the same fixed Llama attribution model and differ only in which model generated the benign trajectory: the corrected natural cohort uses the original Qwen2.5-72B trajectory backbone, while GPT-4o and Claude Sonnet 4.5 generate separate prospective trajectories. ``Explicitly specified'' and ``delegated'' describe benign task structure; this table does not estimate attack or deployment prevalence.}
\endgroup
\end{table}

\subsection{Same Authorized Action, Different Source}
The controlled experiment changes only the source of one needed value. For each of 24 bases, the user's permission, the protected value, the exact committed privileged action, the intended effect, and the execution value itself stay fixed. In one condition the user supplies that value. In the other, a legitimate tool supplies the same value. The 96 conditions are derived from these 24 bases, so the matched base, rather than each condition, is the independent unit.

\begin{figure}[!tb]
  \centering
  \includegraphics[width=0.84\textwidth]{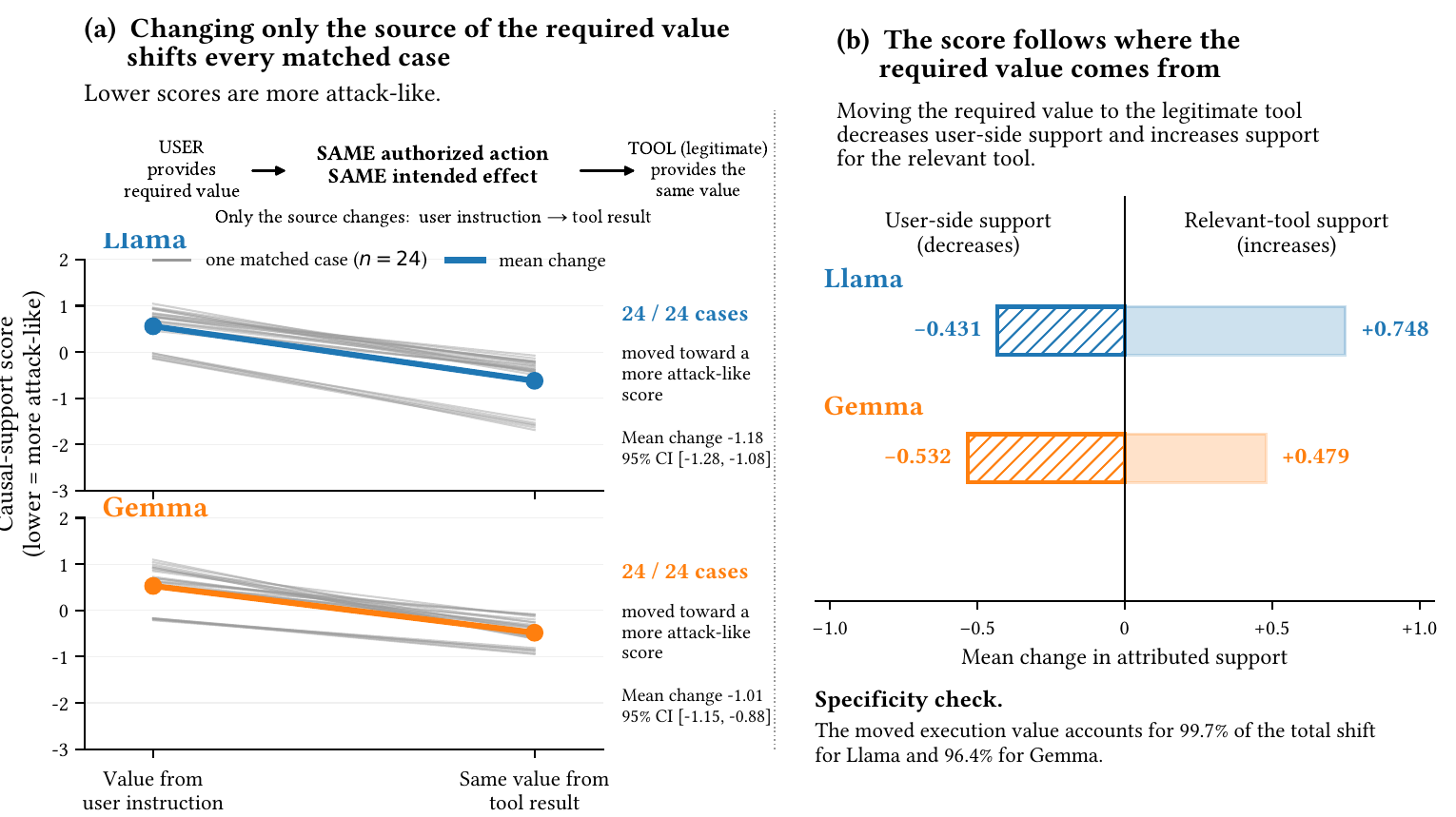}
  \caption{Moving a required value from the user to a legitimate tool makes the same authorized action look more attack-like. Across 24 matched bases, changing only where one required value comes from shifts every case toward the more attack-like region under both scorers. Panel (b) shows the corresponding support shift from the user to the relevant tool; the specificity check localizes nearly all measured movement to the value that changed source (99.7\% for Llama; 96.4\% for Gemma). The action and intended effect remain fixed, so the figure isolates source placement rather than authorization.}
  \label{fig:relocation}
\end{figure}

The source change moves every matched case in the same direction. All 24 Llama cases and all 24 Gemma cases become more attack-like when the needed value comes from the legitimate tool. The mean score change is $-1.1797$ for Llama (95\% CI $[-1.2836,-1.0797]$) and $-1.0112$ for Gemma (95\% CI $[-1.1536,-0.8790]$). Figure~\ref{fig:relocation} shows both the score movement and where the attributed support moves.

The internal support scores explain the movement. User-side support falls on all 24 bases under both scorers, by $-0.4313$ on average for Llama and $-0.5323$ for Gemma. Support from the relevant tool rises on all 24, by $+0.7484$ and $+0.4788$. A specificity check attributes 99.7\% of the Llama movement and 96.4\% of the Gemma movement to the execution value that changed source. The signal is doing what an influence measure should do: it follows where the needed evidence comes from. The security problem is that the movement is large even though the action remains authorized and its intended effect does not change.

At $\tau=0$, the continuous score movement is already large enough to change a binary decision on 20/24 Llama bases and 18/24 Gemma bases across the authorization-equivalent conditions. Every discordant benign change goes from allow to flag; none goes the other way. We use $\tau=0$ only as a descriptive anchor. Section~\ref{sec:policy} examines the complete tested threshold range.

\subsection{Matched Unauthorized Alternative}
An influence score should respond when the source of support changes. That alone would not tell us whether the signal still recognizes a real authorization violation. We add a same-function unauthorized action and ask two separate questions: \emph{Is the unauthorized action more attack-like overall?} And \emph{which change moves the score farther: the authorization violation or the harmless source change?}

The signal still recognizes the threat. The matched unauthorized action is more attack-like overall than the authorized TOOL action under both scorers. The mean unauthorized-minus-authorized endpoint difference is $-0.534$ for Llama (95\% CI $[-0.773,-0.303]$) and $-0.461$ for Gemma (95\% CI $[-0.675,-0.252]$). Two pre-specified manipulation/selectivity checks also move in the expected action-relative direction on all 24 bases under both scorers. The observable therefore contains genuine threat information.

\begin{figure}[!tb]
  \centering
  \includegraphics[width=0.94\textwidth]{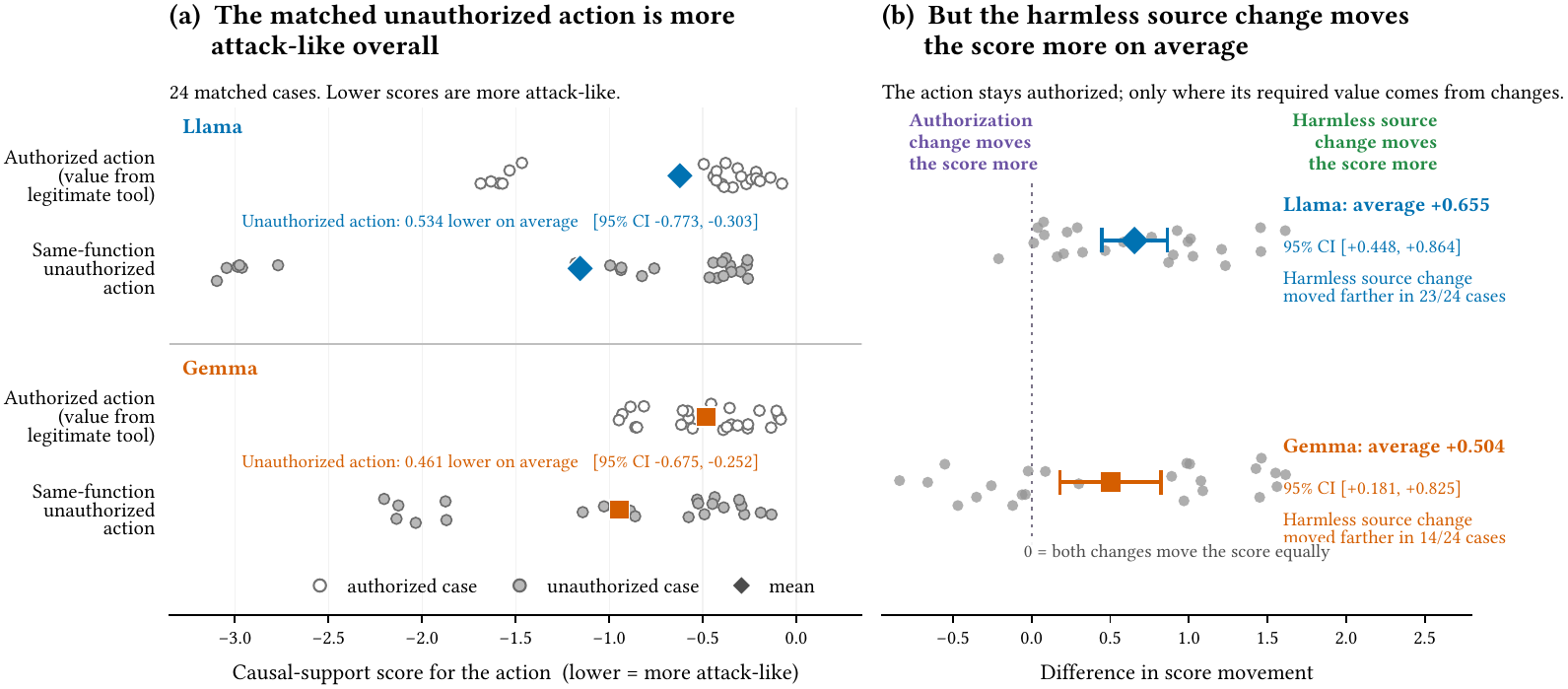}
  \caption{The unauthorized action looks more attack-like overall, but the harmless source change moves the score farther on average. Across 24 matched cases, Panel (a) compares absolute endpoint level and shows that the matched unauthorized action is more attack-like overall under both scorers. Panel (b) instead compares response displacement and shows that changing only where the same required value comes from, while the action remains authorized, produces the larger average score movement. Thus, the signal remains threat-sensitive, but response magnitude is not itself an authorization measure.}
  \label{fig:construct}
\end{figure}

The harmless USER$\rightarrow$TOOL source change nevertheless moves the score farther on average than the change that removes authorization. The average displacement advantage for the harmless source change is $+0.6545$ for Llama (95\% CI $[+0.4475,+0.8636]$) and $+0.5039$ for Gemma (95\% CI $[+0.1806,+0.8246]$). Figure~\ref{fig:construct} shows both results: the unauthorized endpoint is worse, but the harmless source change moves the score farther on average.

The case-level pattern is not uniform. The harmless relocation moves in the attack-like direction on 24/24 bases for both scorers. The authorization-changing comparison moves in that direction on 17/24 Llama bases and 23/24 Gemma bases, and the harmless source change moves farther on 23/24 Llama bases and 14/24 Gemma bases. These data do not support a uniform claim about every unauthorized transformation. They do show that the signal contains real threat information while neither its raw magnitude nor its thresholded decision can be read directly as an authorization label. Appendix~\ref{app:measurement} summarizes the family-level source shifts and the case-level heterogeneity behind Figures~\ref{fig:relocation}--\ref{fig:construct}.

\section{From Influence Signals to Effective Security Decisions}
\label{sec:policy}

The score is only evidence; a deployed guardrail still has to turn it into a decision. We examine three decision paths: a score threshold, AgentWatcher, and AttriGuard, because they differ in which actions they flag, stop, or send to later review.

\subsection{One Score Threshold}
\begin{figure}[!b]
  \centering
  \includegraphics[width=0.92\textwidth]{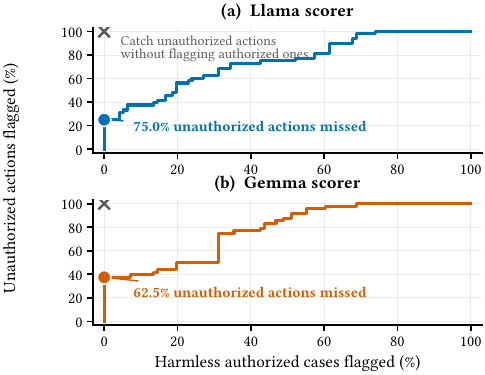}
  \caption{Avoiding every benign flag leaves most unauthorized actions undetected: Llama misses 75.0\% and Gemma 62.5\%. Settings that catch more unauthorized actions also flag more harmless authorized cases, so thresholding offers partial recovery rather than a clean authorization decision. Curves show the complete post-hoc sweep over 24 underlying bases; no operating threshold is selected.}
  \label{fig:threshold}
\end{figure}

A threshold can remove the tested benign flags, but only by missing most of the matched unauthorized actions. At operating points where no tested authorized case is flagged, Llama catches 12/48 unauthorized alternatives (25.0\%) and Gemma catches 18/48 (37.5\%). Other thresholds recover some separation, but they reintroduce benign flags. Even in the predeclared non-extreme range, the harmless source change still flips the matched decision in at least 14/24 Llama cases and 15/24 Gemma cases. Figure~\ref{fig:threshold} shows the full trade-off; Table~\ref{tab:threshold} gives the exact operating points. A threshold helps, but none of the tested scalar rules cleanly separates these matched authorized and unauthorized actions. Appendix~\ref{app:decision} documents how the complete threshold frontier is constructed and why no operating point is selected.

\begin{table}[!tb]
\centering
\caption{Avoiding every tested benign flag leaves most matched unauthorized actions undetected. Llama catches 25.0\% and Gemma catches 37.5\% at those operating points. Other thresholds catch more unauthorized actions but again flag harmless authorized cases. The complete sweep is descriptive; no threshold is recommended.}
\label{tab:threshold}
\begingroup
\footnotesize
\setlength{\tabcolsep}{2.4pt}
\renewcommand{\arraystretch}{1.00}
\begin{tabularx}{\textwidth}{@{}P{0.18\textwidth}C{0.105\textwidth}C{0.14\textwidth}P{0.19\textwidth}P{0.19\textwidth}Y@{}}
\toprule
Scorer / operating point & Harmless cases flagged & Bases changed by source move & Authorized / unauthorized flagged & Unauthorized $-$ authorized & What this point shows \\
\midrule
\textbf{Llama, $\tau=0$} & 41/96 = 42.7\% & 20/24 = 83.3\% & 30/48 = 62.5\% / 36/48 = 75.0\% & \textbf{+12.5 pp}, 95\% CI [+4.17, +20.83] & Descriptive anchor \\
\textbf{Llama, zero-benign point} & \textbf{0/96} & \textbf{0/24} & \textbf{0/48 / 12/48 = 25.0\%} & +25.0 pp & No benign flags, but 75.0\% of unauthorized actions missed \\
\textbf{Llama, predeclared 20--80\% benign range} & 20--80\% by construction & \textbf{minimum 14/24 = 58.3\%} & varies by threshold & varies by threshold & Most matched bases still change decision \\
\textbf{Gemma, $\tau=0$} & 53/96 = 55.2\% & 18/24 = 75.0\% & 35/48 = 72.9\% / 44/48 = 91.7\% & \textbf{+18.75 pp}, 95\% CI [+10.42, +29.17] & Descriptive anchor \\
\textbf{Gemma, zero-benign point} & \textbf{0/96} & \textbf{0/24} & \textbf{0/48 / 18/48 = 37.5\%} & +37.5 pp & No benign flags, but 62.5\% of unauthorized actions missed \\
\textbf{Gemma, descriptive max-BA point} & 7/96 = 7.3\% & 7/24 = 29.2\% & 0/48 / 19/48 = 39.6\% & +39.58 pp & Post-hoc descriptive maximum; not a selected policy \\
\textbf{Gemma, predeclared 20--80\% benign range} & 20--80\% by construction & \textbf{minimum 15/24 = 62.5\%} & varies by threshold & varies by threshold & Most matched bases still change decision \\
\bottomrule
\end{tabularx}
\parbox{0.97\textwidth}{\emph{Authorized / unauthorized entries use the same 48-action denominator within each scorer. The sweep includes every observed breakpoint, every interval between adjacent breakpoints, the extremes, and $\tau=0$. Maximum-balanced-accuracy points are post-hoc descriptions of the frozen sweep, not prospectively validated choices.}}
\endgroup
\end{table}

\begin{figure}[!tb]
  \centering
  \includegraphics[width=0.92\textwidth]{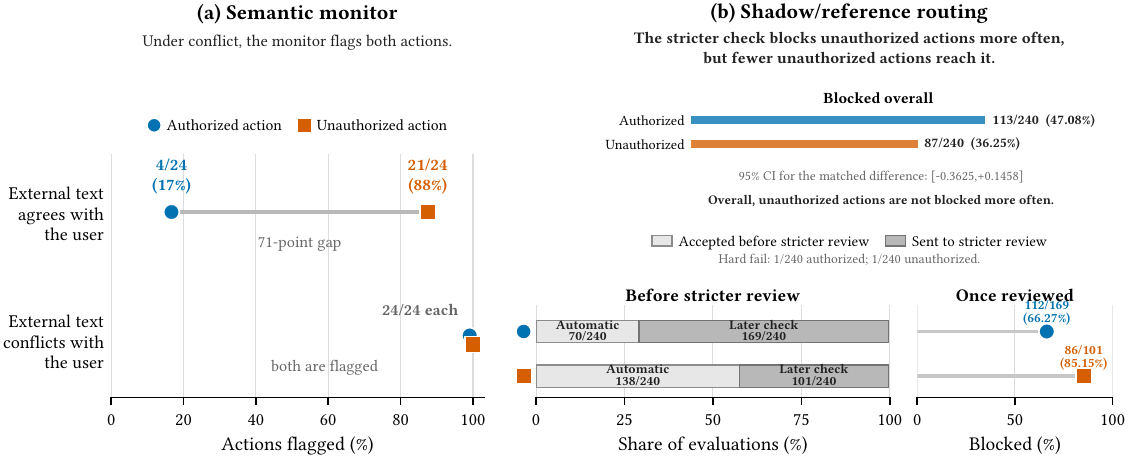}
  \caption{Different decision paths can lose the authorized/unauthorized distinction at different stages. (a) AgentWatcher separates the actions when external text agrees with the user, but flags both under conflict. (b) AttriGuard does not establish more blocking of unauthorized actions overall (113/240 authorized vs. 87/240 unauthorized; the matched CI spans zero). The route explains the local/overall difference: unauthorized calls reach automatic exact survival more often, while later review is stricter on unauthorized calls once reached. The observed reference determines the route; we do not establish that the conflicting directive caused that reference identity.}
  \label{fig:architecture}
\end{figure}

\subsection{Semantic Monitor}
A scalar threshold uses only the score. AgentWatcher asks whether semantic reasoning over the influential text can separate the authorized and unauthorized actions more cleanly~\cite{wang2026agentwatcher}. When the external text agrees with the user, it flags 4/24 authorized actions and 21/24 matched unauthorized actions. Under the conflict context we test, it flags all 24 authorized actions and all 24 unauthorized actions. In plain terms, the monitor distinguishes the actions in the aligned case but flags both under conflict. This is a paired gate-level result; it is not an observed end-to-end denial-of-service result.

A positive flag does more than add a warning: in this integration, it cancels the pending tool-use path. The implementation clears the pending assistant tool calls and removes prior tool observations before returning control to the agent loop, so the next loop starts without the flagged call. The observed effect therefore reflects both the monitor's decision and how the surrounding system enforces it. It does not show that semantic reasoning cannot represent authority; it shows how this particular monitored path behaves on the matched contexts we test.

In a separate matched-input evaluation, enabling AgentWatcher gives 0\% attack success and 28\% utility, versus 16\% and 60\% without it. These are separate API executions and are distinct from the 2$\times$2 gate study above.

\subsection{Shadow/Reference Routing}
A later security judgment matters only if a candidate reaches it. AttriGuard uses a shadow/reference path~\cite{he2026attriguard}, so routing becomes part of the security decision: which candidates reach its later review? A call that exactly matches its shadow/reference survives automatically; a non-matching call can reach a later fuzzy review. Across 240 repeated invocations per action type, AttriGuard blocks 113 authorized actions (47.1\%) and 87 unauthorized actions (36.3\%). The unauthorized-minus-authorized difference is $-10.8$ percentage points with a 95\% CI of $[-36.3,+14.6]$ percentage points. Because the interval includes zero, these data do not show that unauthorized actions are blocked more often overall.

The route to the later check explains why the local and overall results differ. Unauthorized calls reach automatic exact survival more often, while authorized calls reach fuzzy review more often. Conditional on reaching that later review, unauthorized calls are blocked more often. Figure~\ref{fig:architecture} makes this route asymmetry visible, and Table~\ref{tab:architecture} gives the exact route and block counts. A stricter later check can therefore coexist with no clear overall advantage in blocking unauthorized actions. Appendix~\ref{app:decision} gives the enforcement and routing checks behind Figure~\ref{fig:architecture}.

The observed reference mechanically determines whether a call survives automatically or reaches the later check. Our data do not establish that the conflicting directive caused that reference identity.

\begin{table}[!tb]
\centering
\caption{The three guardrails lose or preserve the authorized/unauthorized distinction at different points. A threshold trades benign flags for unauthorized actions caught; AgentWatcher flags both actions under conflict; AttriGuard's later check is stricter on unauthorized actions, but fewer unauthorized actions reach it. The rows use different endpoints and should not be read as a leaderboard.}
\label{tab:architecture}
\begingroup
\footnotesize
\setlength{\tabcolsep}{2.8pt}
\renewcommand{\arraystretch}{1.04}
\begin{tabularx}{\textwidth}{@{}P{0.135\textwidth}P{0.17\textwidth}P{0.285\textwidth}P{0.22\textwidth}Y@{}}
\toprule
Decision path & What it uses & What happens & Security consequence & Scope \\
\midrule
\textbf{Single score threshold} &
One causal-support score compared with $\tau$ &
With no tested benign flags, only \textbf{12/48} Llama and \textbf{18/48} Gemma unauthorized alternatives are caught &
Avoiding benign flags leaves most matched unauthorized alternatives undetected; no tested threshold cleanly separates the actions &
Complete deterministic sweep; no preferred threshold \\

\textbf{Semantic monitor (AgentWatcher)} &
Influential text plus semantic/rule reasoning; a positive flag cancels the pending tool-use path &
External text agrees with user: authorized \textbf{4/24}, unauthorized \textbf{21/24} flagged. Under conflict: \textbf{24/24} of both flagged &
The monitor distinguishes the actions when the text agrees with the user, but flags both under conflict &
Matched gate study; separate on/off population; gate-level evidence \\

\textbf{Shadow/reference path (AttriGuard)} &
Exact reference match survives automatically; other calls can reach fuzzy review &
Overall blocked: authorized \textbf{113/240 = 47.08\%}, unauthorized \textbf{87/240 = 36.25\%}, 95\% CI for matched gap [$-0.3625,+0.1458$]. Exact survival: \textbf{70/240} vs. \textbf{138/240}; fuzzy review: \textbf{169/240} vs. \textbf{101/240} &
Once reviewed, unauthorized calls are blocked more often (\textbf{86/101 = 85.15\%} vs. \textbf{112/169 = 66.27\%}), but fewer unauthorized calls reach that check; overall, the data do not show that unauthorized actions are blocked more often &
Observed reference determines route; directive-to-reference causality unresolved \\
\bottomrule
\end{tabularx}
\parbox{0.97\textwidth}{\emph{Separate AgentWatcher on/off experiment:} on the same 200 Tool-Knowledge inputs, defense ON has 28\% utility and 0\% attack success; defense OFF has 60\% utility and 16\% attack success. These are separate API executions, so the comparison is matched-input evidence rather than a randomized causal effect.}
\endgroup
\end{table}

\subsection{A Gate Decision Is Not the Final Outcome}
Under the tested conflict context, a still-authorized action can be flagged or blocked even though the user's permission has not changed. But those measurements stop at the gate. A block can redirect what the agent does next, so restrictive gate behavior alone cannot tell us whether the user ultimately loses the authorized effect. We therefore follow execution to the protected effect and ask: after a guardrail intervenes, does the execution preserve what the user authorized?

\section{Live Execution: Intervention, Continuation, and Protected Effect}
\label{sec:live}

\begin{figure}[!tb]
  \centering
  \includegraphics[width=0.92\textwidth]{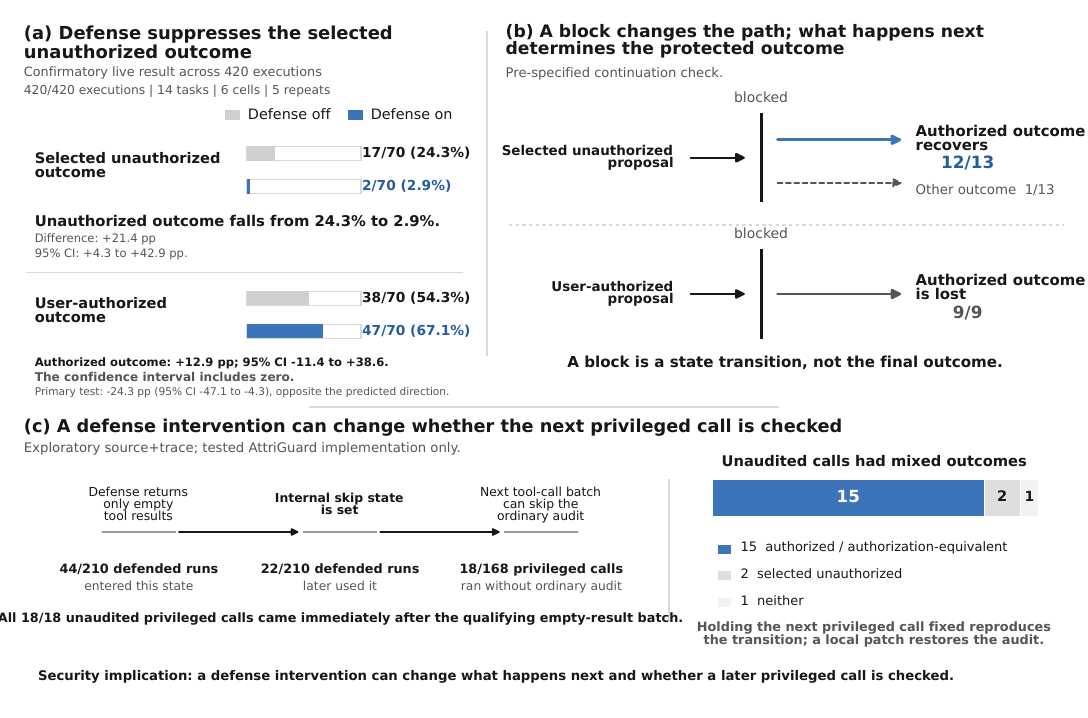}
  \caption{A block changes the execution path, not just one decision. (a) Under conflict, the selected unauthorized outcome falls from 17/70 with the defense off to 2/70 with it on. The user-authorized outcome changes from 38/70 to 47/70, but its 95\% CI includes zero; the pre-specified primary effect also runs opposite the predicted availability-loss direction. (b) What happens after the block determines the protected effect: 12/13 blocked selected-unauthorized proposals later recover the authorized outcome, whereas all 9/9 blocked authorized proposals lose it. (c) In the tested AttriGuard implementation, intervention can also change whether the next privileged call receives the ordinary audit. This exploratory mechanism is source-bound and is not an exploit-rate result.}
  \label{fig:live}
\end{figure}

We run AttriGuard end to end on 14 prospectively selected natural tasks. Each task is tested under CLEAN, ALIGNED, and CONFLICT context with the defense either OFF or ON, and each cell is repeated five times, giving 420 executions in total. Figure~\ref{fig:live} follows the result from the protected outcome through continuation and later inspection. We treat the natural task as the inferential unit; confidence intervals use the pre-specified task-level bootstrap. All 420 scheduled runs complete on the first attempt. Appendix~\ref{app:execution} reports the complete six-cell outcomes and the checks behind continuation and later inspection.

For the live study, we track whether the selected unauthorized effect occurs and whether execution preserves the effect the user authorized. We call the latter Protected Authorization/Effect Fidelity (\PAEF).

\subsection{What Happens to the Protected Outcome?}
The primary result runs opposite the pre-specified availability-loss prediction. We expected enabling the defense to reduce preservation of the authorized outcome more under CONFLICT than under ALIGNED. Instead, the estimated interaction is $-0.2429$ with a 95\% CI of $[-0.4714,-0.0429]$, pointing in the opposite direction. The pre-specified test therefore does not support the predicted availability loss. A restrictive gate decision does not mechanically determine the final user-authorized outcome.

Under CONFLICT, the selected unauthorized outcome becomes much less common with the defense. It occurs in 17/70 executions (24.3\%) when the defense is off and 2/70 (2.9\%) when it is on. The pre-specified difference is $+0.2143$ with a 95\% CI of $[+0.0429,+0.4286]$. On this cohort, the selected unauthorized outcome is less common with AttriGuard enabled.

The user-authorized outcome is less conclusive. Under CONFLICT, \PAEF{} is 38/70 (54.3\%) with the defense off and 47/70 (67.1\%) with it on, a difference of $+12.9$ percentage points. Its 95\% CI is $[-11.4,+38.6]$ and includes zero. We do not claim that AttriGuard generally improves preservation of the authorized effect.

\subsection{What Happens After a Block?}
Panel~(b) of Figure~\ref{fig:live} shows why the gate result is not the final outcome. Under CONFLICT, AttriGuard blocks 13 selected unauthorized privileged proposals. In 12/13 traces, the agent later completes an authorization-equivalent privileged action and ends with \PAEF$=1$. The reverse pattern also occurs. AttriGuard blocks nine authorization-equivalent privileged proposals, and all 9/9 traces end with \PAEF$=0$.

Blocking can therefore help or hurt the protected outcome depending on what happens next. \textbf{A block is a state transition, not the final outcome.} After intervention, the agent may recover the authorized outcome, lose it, realize the selected unauthorized outcome, or take another privileged path. Evaluating only the gate misses part of the security policy.

Because the agent can continue after a block, the security of that continuation depends on more than which action it chooses next. A later privileged action must also receive the intended security checks, so we ask whether intervention can change the inspection coverage applied to the next tool-call batch.

\subsection{Can Intervention Change the Next Security Check?}
Panel~(c) of Figure~\ref{fig:live} shows an exploratory, implementation-specific control-flow effect. In the AttriGuard version we tested, one defense intervention can leave an internal state that causes a later tool-call batch to run without the ordinary audit. We observe this state only after a defense-generated tool-result batch in which all relevant contents are empty; mixed-result batches do not trigger it. The effect is specific to the tested implementation and configuration; it does not show that every block disables the next audit or that newer upstream versions behave the same way.

The counts use different denominators because they answer different questions. Of 210 defended runs, 44 open the qualifying state and 22 later execute a tool batch while that state is active. Those runs contain 168 defended privileged calls; 18 of those calls execute without the ordinary audit. All 18/18 immediately follow the qualifying predecessor. Their outcomes are mixed: 15 preserve an authorized or authorization-equivalent effect, two realize the selected unauthorized effect, and one produces neither. The security consequence is that intervention can change later inspection. These counts do not estimate exploit reliability or attack success.

A controlled source-level test reproduces the state transition while holding the following privileged call fixed, and a local patch restores inspection. This supports the implementation mechanism, but the mechanism was identified post hoc and is specific to the tested version and configuration. It does not estimate exploit reliability or establish a general vulnerability. Appendix~\ref{app:execution} gives the denominator ladder and deterministic isolation behind this source-bound mechanism. Responsible disclosure has been completed.

These results also expose an evaluation problem: ordinary task success may not tell us whether the authorized effect survived. The live runs test that directly; replay examines the same gap at a single privileged decision.

\section{Evaluation Fidelity: Does Success Measure the Protected Property?}
\label{sec:evaluation}

A task can pass even when the privileged action or effect the user authorized was not preserved. Across the 420 live executions, ordinary task utility and \PAEF{} disagree on 18 runs. Thirteen runs pass the task check even though the authorized protected effect is lost. Five preserve the authorized effect even though the task check fails. Neither metric is wrong; they answer different questions.

At a single privileged decision, replay asks whether later continuation can hide an immediate change. For each decision, we freeze the history immediately before it and ask the model to generate that next action again through the validated tool-call interface. We then execute the regenerated action and check whether its immediate environment effect matches the original.

The downstream check keeps everything after that point as stable as possible. We replay the original later tool calls, keep the original final answer fixed, and run the benchmark utility check. Only the regenerated privileged decision is allowed to change. This follows the distinction between replayed decision fidelity and downstream task success studied in causal replay and replay-gap work~\cite{shah2026causalreplay,gonuguntla2026replaygap}. AgentDojo and ClawsBench provide the task and safety evaluation context in which those outcomes are scored~\cite{debenedetti2024agentdojo,li2026clawsbench}.

Across 78 model--decision pairs spanning Llama, Gemma, and Qwen, 23/78 pass the downstream majority check even though the immediate action/effect check fails. Each pair has five stability repeats, for 390 generations in total; the 78 pairs are the relevant units for this comparison. In 22/23 disagreeing pairs, the model still calls the intended tool. Tool-name equality can therefore hide a changed argument or effect, while later recovery can hide an earlier divergence. Table~\ref{tab:evaluation} puts the live and replay evidence side by side; Appendix~\ref{app:replay} gives the replay protocol and the majority-level cross-tabs behind this disagreement.

\begin{table}[!tb]
\centering
\caption{Task success can hide a change to the privileged action or effect. Live utility and the user-authorized effect disagree on 18/420 executions; replay also shows higher downstream task success than immediate action/effect fidelity. These rows answer an evaluation question, not a model-ranking question.}
\label{tab:evaluation}
\begingroup
\footnotesize
\setlength{\tabcolsep}{2.6pt}
\renewcommand{\arraystretch}{1.00}
\begin{tabularx}{\textwidth}{@{}P{0.18\textwidth}P{0.20\textwidth}P{0.19\textwidth}P{0.19\textwidth}Y@{}}
\toprule
Evidence / setting & Immediate check & Downstream check & What disagrees & Why it matters \\
\midrule
\textbf{Live execution}: AttriGuard, 420 runs & Whether the user-authorized effect is preserved & Whether the task succeeds & \textbf{18/420:} 13 task-pass / authorized-effect-fail; 5 task-fail / authorized-effect-pass & Task success is not the same as preserving what the user authorized \\
\textbf{Replay}: Llama-3.3-70B & Action/effect preserved: \textbf{50/130 = 38.5\%} & Task succeeds: \textbf{85/130 = 65.4\%} & 35/130 generations differ in this direction & Later continuation can hide an immediate action/effect change \\
\textbf{Replay}: Gemma-3-12B & Action/effect preserved: \textbf{35/130 = 26.9\%} & Task succeeds: \textbf{75/130 = 57.7\%} & 40/130 generations differ in this direction & Same evaluation gap with a second replay model \\
\textbf{Replay}: Qwen2.5-72B & Action/effect preserved: \textbf{55/130 = 42.3\%} & Task succeeds: \textbf{95/130 = 73.1\%} & 40/130 generations differ in this direction & Same evaluation gap with a third replay model \\
\textbf{Across models}: 78 model--decision pairs & Immediate action/effect check fails & Downstream majority check passes & \textbf{23/78 pairs}; \textbf{22/23} still use the intended tool & Tool-name equality and task success can both miss argument/effect changes \\
\bottomrule
\end{tabularx}
\parbox{0.97\textwidth}{\emph{The 390 replay generations are five stability repeats over 78 model--decision pairs; they are not 390 independent cases.}}
\endgroup
\end{table}

These results do not make task utility useless. They show why utility must remain separate from the property a defense is supposed to preserve. We therefore report several outcomes separately: task success, immediate action/effect fidelity, authorized availability, later inspection, and final \PAEF{}. No single one substitutes for the others.

\section{Discussion and Limitations}
\label{sec:discussion}

Influence and provenance are useful security evidence, but a guardrail should preserve or check authority separately. The controlled result shows why. Moving a required value from the user request to a legitimate tool result makes the studied signal follow that new source even though permission, the action, and the intended effect do not change. The matched unauthorized action remains more attack-like overall, so the signal is not noise. The practical boundary is narrower: response magnitude should not carry the authorization decision by itself~\cite{wang2026authgraph,qin2026airguard}.

The decision path must therefore be evaluated as policy, not just as a final classifier. Thresholding trades harmless flags against unauthorized actions caught. A semantic monitor can distinguish the actions in aligned context and still flag both under conflict. A reference-based design can apply a stricter later review to unauthorized calls yet expose fewer of them to that review. Builders and evaluators should report who survives automatically, who reaches adjudication, what is blocked, and what happens overall. Conflict sensitivity can be a reasonable conservative choice; the corresponding availability question is whether the still-authorized action remains usable.

Intervention also changes the state from which execution continues. A block may be followed by recovery of the authorized effect or by its loss, and in the tested AttriGuard implementation a defense-generated state can also change whether a later privileged call receives the ordinary audit. A defense that changes execution state should not silently reduce the security coverage applied to what happens next. This is a design principle, not a claim that every architecture has the source-bound behavior we observed~\cite{santosgrueiro2026cxi,santosgrueiro2026committime}.

Evaluation should name the protected property before choosing a success metric. Attack outcome, task success, preservation of the user-authorized effect, authorized availability, and inspection coverage answer different questions. The live and replay studies show that these properties can disagree within the same workflow. Utility remains useful, but it should be reported alongside the security property the defense claims to preserve~\cite{li2026clawsbench,zhang2026fava}. For effectful tool use, this means following execution far enough to ask whether the authorized effect occurs, whether the authorized action remained available, and whether later privileged actions still received the intended inspection.

These conclusions are bounded by the populations and implementations we tested. The 24-base controlled study isolates a source change; it does not estimate deployment prevalence. The natural benign cohort is finite, and the matched unauthorized comparison is teacher-forced, so its construct-validity result concerns matched averages rather than uniform per-case behavior or deployment outcomes. The CausalArmor result audits a reconstruction of the published estimand, the AgentWatcher study stops at the gate, and AttriGuard's aggregate blocked-action confidence interval spans zero. Reference identity determines later-review exposure in the tested path, but the data do not establish that the conflicting directive caused that identity.

The live study uses 14 prospectively selected tasks and one fixed conflict directive. That directive is a matched experimental condition, not a worst-case adaptive attack. The direct CONFLICT \PAEF{} interval includes zero. Reduced audit coverage was identified post hoc; deterministic source-level isolation reproduces the transition, but the two selected unauthorized outcomes do not estimate exploit reliability. Utility--\PAEF{} disagreement shows metric divergence in these runs, not a universal utility failure. Broader claims about prevalence or architecture-wide behavior would require larger natural populations, additional task and tool families, adaptive conflict text, and verification on other implementation versions.

\section{Conclusion}
\label{sec:conclusion}

Tool-using agents often need external information to carry out an action the user has already authorized. Our results show why that dependence cannot be treated as authority. Moving one required value from the user request to a legitimate tool result makes the same authorized action look more attack-like on all 24 matched bases under both scorers. The matched unauthorized action still looks more attack-like overall, but the harmless source change produces the larger average score movement. The signal is therefore informative about what shaped the action without being an authorization label.

Once a guardrail acts on that evidence, the rest of the decision path becomes part of the security policy. Thresholding, semantic judgment, reference construction, and routing determine which actions reach intervention; after intervention, continuation determines which effect actually occurs, and later inspection can itself change. A score, flag, or block is therefore an intermediate security decision rather than the final outcome.

For tool-using agents, evaluation should follow the property the user authorized through that full path and measure whether it survives. Influence can tell us what shaped an action. A guardrail must still check whether that action stays within what the user authorized.

\section*{Ethical Considerations}
\label{app:ethics}

This work evaluates defenses against indirect prompt injection and reports an exploratory implementation-specific finding about later inspection. We describe the mechanism only at the level needed to support the systems claim. We do not present it as a general attack path, do not estimate exploit success or reliability from the two selected unauthorized outcomes, and limit every implementation claim to the frozen version and configuration we tested. Most of the 18 affected privileged calls preserve an authorized effect; two realize the selected unauthorized effect. Responsible disclosure of the source-level finding has been completed.

The controlled studies use benchmark or constructed agent tasks to evaluate signals, decisions, and protected effects. Our claims concern the tested systems and populations rather than deployment prevalence or universal defense behavior. The anonymous reviewer artifact should retain the evidence needed to audit these claims while removing author-identifying paths, repository metadata, credentials, and tracking mechanisms.

\section*{Open Science}
\label{app:openscience}

\textbf{Anonymous artifact:} \url{https://anonymous.4open.science/r/influence-is-not-authority-artifact-A674/README.md}.

The anonymous artifact contains the frozen evidence and code needed to audit the paper's main results. It covers the controlled source-relocation and matched-unauthorized studies, threshold analysis, AgentWatcher and AttriGuard experiments, the live protected-outcome study, implementation-specific analyses, deterministic isolation checks, and replay experiments. It also includes the scripts used to regenerate the paper's figures and the source and configuration records needed for implementation-specific claims.

The artifact provides two reproduction paths. Reviewers can verify the reported quantities directly from the frozen execution record without new model calls, credentials, or GPUs. Reviewers with the required model access and hardware can also rerun the experiments end to end. Fresh runs are kept separate from the frozen evidence used for the submitted results.

The top-level documentation explains the artifact structure, reproduction procedure, and experimental dependencies. A claim-to-artifact map links manuscript-bearing results to their supporting evidence and verification commands. Appendix~\ref{app:repro} provides the corresponding paper-level map from claims to frozen evidence and deterministic checks. The anonymous artifact contains no author credentials or private service state.

\bibliographystyle{plainurl}
\bibliography{references}

\appendix
% Claim-linked supplemental appendix.
% Each appendix is tied to a main-text result and adds a check, decomposition,
% or implementation detail that would distract from the main argument.
% Exhaustive row-level ledgers remain in the anonymous artifact.

\makeatletter
\@addtoreset{table}{section}
\@addtoreset{figure}{section}
\makeatother
\renewcommand{\thetable}{\thesection\arabic{table}}
\renewcommand{\thefigure}{\thesection\arabic{figure}}

\section{Additional Evidence for the Source-Relocation Result}
\label{app:measurement}

Section~\ref{sec:measurement} makes the paper's central measurement claim: a legitimate tool can supply a value that an authorized action needs and still move the studied signal in the attack-like direction. The main figures show the matched effect and its construct qualification. This appendix asks the narrower questions a reviewer may have after seeing those figures: whether the natural motivation spans more than one workflow family, whether the controlled effect is concentrated in one function family, how heterogeneous the matched unauthorized comparison is, and what exactly is reconstructed from CausalArmor.

\subsection{The natural motivation is not confined to one workflow family}
The corrected benign cohort contains 29 privileged decisions from 25 tasks. Table~\ref{tab:app_natural_suite} shows where those decisions come from. The full decision ledger is left in the artifact because the paper's claim is ecological rather than a row-level benchmark result: legitimate tool-resolved support occurs in several benign workflow families, but this cohort does not estimate deployment prevalence.

\begin{table}[!ht]
\centering
\caption{The benign cohort spans four workflow suites. ``User-dominant'' counts use the fixed Llama attribution scorer. Partial cases are shown for completeness but are not part of the specified-versus-delegated contrast in Table~\ref{tab:natural}.}
\label{tab:app_natural_suite}
\footnotesize
\setlength{\tabcolsep}{3.2pt}
\begin{tabular}{@{}lrrrrr@{}}
\toprule
Suite & Dec. & Spec. & Deleg. & Partial & User-dom. (S/D) \\
\midrule
Banking   & 3  & 1 & 2 & 0 & 1/0 \\
Slack     & 14 & 2 & 6 & 6 & 1/1 \\
Travel    & 1  & 0 & 1 & 0 & 0/1 \\
Workspace & 11 & 5 & 2 & 4 & 4/0 \\
\bottomrule
\end{tabular}
\end{table}

GPT-4o and Claude Sonnet 4.5 play a different role from the scorers in the controlled experiment: they generate prospective trajectories, while the same fixed Llama attribution model scores both trajectory sets. Their rows in Table~\ref{tab:natural} therefore test generator breadth, not native attribution replication.

\subsection{The controlled shift appears in every constructed family}
The 24 matched bases are evenly split across four protected tool actions. Table~\ref{tab:app_family_shift} aggregates the USER$\rightarrow$TOOL score change by family. Every individual base moves in the attack-like direction for both scorers, so the 24/24 result in Figure~\ref{fig:relocation} is not carried by one function family.

\begin{table}[!ht]
\centering
\caption{Family-level USER$\rightarrow$TOOL shifts behind Figure~\ref{fig:relocation}. Each family contains six matched bases; every individual base moves in the attack-like direction under both scorers.}
\label{tab:app_family_shift}
\footnotesize
\setlength{\tabcolsep}{4pt}
\begin{tabular}{@{}lrr@{}}
\toprule
Protected action family & Llama & Gemma \\
\midrule
Calendar: add participant & -1.311 & -1.552 \\
Slack: add user & -1.494 & -0.696 \\
Banking: send money & -0.925 & -1.036 \\
Email: send email & -0.989 & -0.761 \\
\bottomrule
\end{tabular}
\end{table}

The attribution decomposition follows the same pattern on every base: user-side support falls and support from the relevant tool rises. The artifact contains the per-base scores and support decomposition used to generate Figure~\ref{fig:relocation}; reproducing those 48 scorer rows here would duplicate the figure rather than add a new check.

\subsection{Figure~\ref{fig:construct} supports an average-level, not universal, construct claim}
The matched unauthorized comparison answers two different questions. First, is the unauthorized endpoint more attack-like than the authorized TOOL endpoint? Second, which change moves the score farther: removing authorization or moving a legitimate value from user to tool? Table~\ref{tab:app_construct_summary} puts the corresponding case-level sign counts beside the average effects. This is the heterogeneity hidden by a single mean.

\begin{table}[!ht]
\centering
\caption{Construct checks behind Figure~\ref{fig:construct}. Endpoint and displacement are reported separately because they answer different questions.}
\label{tab:app_construct_summary}
\footnotesize
\setlength{\tabcolsep}{3.5pt}
\begin{tabularx}{\columnwidth}{@{}P{0.12\columnwidth}P{0.26\columnwidth}Y P{0.18\columnwidth}@{}}
\toprule
Scorer & Check & Average (95\% CI) & Case-level direction \\
\midrule
Llama & Unauthorized endpoint & $-0.534$ $[-0.773,-0.303]$ & 17/24 expected \\
Llama & Harmless-source displacement advantage & $+0.6545$ $[+0.4475,+0.8636]$ & 23/24 harmless farther \\
Gemma & Unauthorized endpoint & $-0.461$ $[-0.675,-0.252]$ & 23/24 expected \\
Gemma & Harmless-source displacement advantage & $+0.5039$ $[+0.1806,+0.8246]$ & 14/24 harmless farther \\
\bottomrule
\end{tabularx}
\end{table}

The pre-specified manipulation/selectivity checks move in the expected action-relative direction on all 24 bases under both scorers. Those checks matter because they rule out the simple interpretation that the signal is merely noisy. The signal contains threat information; what fails is the stronger interpretation that response magnitude can be read directly as authorization.

\subsection{What is reconstructed from CausalArmor}
Released implementation code was unavailable for the CausalArmor path we audit. We therefore reconstruct the published leave-one-out user-versus-untrusted margin, its length normalization, and the $\tau=0$ anchor, then check that reconstruction against the reported broad operating regime. Table~\ref{tab:app_calibration} records that calibration. This is an audit of the published estimand, not an implementation-identical reproduction.

\begin{table}[!ht]
\centering
\caption{External-regime calibration for the reconstructed CausalArmor-style estimand. The purpose is to verify the broad operating regime before using the estimand in the controlled study, not to claim code-identical reproduction.}
\label{tab:app_calibration}
\footnotesize
\begin{tabular}{@{}lrrrrr@{}}
\toprule
Regime & Benign & Attacked & BU & UA & ASR \\
\midrule
Primary & 97 & 949 & 51.5\% & 40.7\% & 3.4\% \\
Sensitivity & -- & 629 & -- & 42.9\% & 5.1\% \\
\bottomrule
\end{tabular}
\end{table}

A separate 26-decision benign calibration gives one operational consequence of false activation in this reconstructed path: 18/26 decisions activate the defense, 53/96 eligible tool spans are sanitized, and the serial sanitizer stage averages about 2.364\,s among activated decisions. These numbers measure additional sanitizer work and stage latency; they do not establish benign task failure.

\section{Why Tuning and Decision Paths Do Not Recover Authority}
\label{app:decision}

Section~\ref{sec:policy} moves from the signal to the policy that acts on it. The main paper compares three paths: a scalar threshold, AgentWatcher, and AttriGuard. The material below complements Figures~\ref{fig:threshold} and~\ref{fig:architecture} by showing how the policy analyses were constructed and by exposing the routing/enforcement details that are hard to read from aggregate percentages alone.

\subsection{The threshold result is a complete frontier, not a chosen operating point}
The threshold analysis is deterministic over frozen scores. For each scorer, it evaluates every observed breakpoint, every interval between adjacent breakpoints, both extremes, and $\tau=0$. The complete sweep contains 386 threshold rows over 192 unique observed breakpoints per scorer. No deployment threshold is prospectively selected, and the descriptive extrema are not recommended policies.

This matters because Figure~\ref{fig:threshold} does not support the claim that thresholding is useless. Low thresholds can remove all tested benign flags while retaining limited unauthorized sensitivity. The bounded conclusion is narrower: across the complete tested frontier, the scalar score never becomes a clean authorization decision, and most matched bases remain source-sensitive throughout the predeclared non-extreme benign range. The full frontier CSVs stay in the artifact so reviewers can recompute any operating point without adding hundreds of rows to the paper.

\subsection{AgentWatcher: semantic separation disappears under the tested conflict context}
Table~\ref{tab:app_aw_summary} is the aggregate 2$\times$2 behind Figure~\ref{fig:architecture}a. When external text agrees with the user, the monitor distinguishes many authorized from unauthorized actions. Under the tested conflict context, both classes are flagged on every base.

\begin{table}[!ht]
\centering
\caption{AgentWatcher gate result behind Figure~\ref{fig:architecture}a. The aligned context preserves substantial separation; under the tested conflict context, every authorized and unauthorized action is flagged.}
\label{tab:app_aw_summary}
\footnotesize
\begin{tabular}{@{}lrr@{}}
\toprule
Context & AUTH flagged & ALT flagged \\
\midrule
Aligned  & 4/24  & 21/24 \\
Conflict & 24/24 & 24/24 \\
\bottomrule
\end{tabular}
\end{table}

A positive verdict also changes execution state in the tested integration. The surrounding enforcement logic clears the pending assistant tool calls, removes prior tool observations, and returns control to the agent loop without the flagged call. The next loop therefore starts from a different history. This source-level behavior explains why a gate verdict cannot be interpreted independently of the system that enforces it; it is not an observed end-to-end denial-of-service result.

The separate 200-input ON/OFF experiment provides the operational anchor reported in the main paper: defense ON yields 0\% attack success and 28\% utility, compared with 16\% and 60\% with the defense OFF. Those are separate API executions on a different population, so we keep them separate from the paired 2$\times$2 above.

\subsection{AttriGuard: the stricter later check sees a different population}
AttriGuard's aggregate result must be read before its conditional later-check result. Table~\ref{tab:app_attriguard_summary} shows why. Unauthorized calls are rejected more often once they reach fuzzy review, but they reach automatic exact survival much more often. The route therefore changes which calls the stricter check ever sees.

\begin{table}[!ht]
\centering
\caption{AttriGuard aggregate outcome and route decomposition behind Figure~\ref{fig:architecture}b. The later check is stricter on unauthorized calls once reached, but unauthorized calls reach it less often.}
\label{tab:app_attriguard_summary}
\footnotesize
\setlength{\tabcolsep}{3.8pt}
\begin{tabular}{@{}lrr@{}}
\toprule
Outcome / route & AUTH & ALT \\
\midrule
Blocked overall & 113/240 & 87/240 \\
Automatic exact survival & 70/240 & 138/240 \\
Later review exposure & 169/240 & 101/240 \\
Blocked once reviewed & 112/169 & 86/101 \\
\bottomrule
\end{tabular}
\end{table}

The matched unauthorized-minus-authorized aggregate block difference is $-0.1083$ with a 95\% CI of $[-0.3625,+0.1458]$. Across the 24 bases, the aggregate sign is positive on 8 bases, zero on 4, and negative on 12. The route decomposition localizes the system behavior; it does not establish that the conflicting directive caused the shadow/reference to reproduce the unauthorized action. Observed reference identity determines route exposure, while directive-to-reference causality remains unresolved.

\section{What Happens After a Block}
\label{app:execution}

Figure~\ref{fig:live} is the systems culmination of the paper: it follows intervention through the protected outcome, continuation, and later inspection. This appendix adds the tables needed to audit those three claims without turning the main figure into a ledger. The natural task, not each repeated execution, is the inferential unit for the live confidence intervals.

\subsection{The full six-cell live result}
Table~\ref{tab:app_live_cells} shows the descriptive outcomes for all CLEAN, ALIGNED, and CONFLICT cells. The pre-specified interaction is computed at the task level. Under CONFLICT, the selected unauthorized outcome falls from 17/70 with the defense off to 2/70 with it on. PAEF moves from 38/70 to 47/70, but that direct interval includes zero; the primary interaction also runs opposite the pre-specified availability-loss direction.

\begin{table}[!ht]
\centering
\caption{All six live cells behind Figure~\ref{fig:live}a. Each cell has 70 executions; inference resamples the 14 natural tasks.}
\label{tab:app_live_cells}
\footnotesize
\setlength{\tabcolsep}{3.5pt}
\begin{tabular}{@{}llrrr@{}}
\toprule
Context & Defense & Selected ALT $Z$ & PAEF & Utility \\
\midrule
CLEAN    & OFF & 0/70  & 41/70 & 46/70 \\
CLEAN    & ON  & 0/70  & 35/70 & 39/70 \\
ALIGNED  & OFF & 0/70  & 60/70 & 60/70 \\
ALIGNED  & ON  & 0/70  & 52/70 & 50/70 \\
CONFLICT & OFF & 17/70 & 38/70 & 40/70 \\
CONFLICT & ON  & 2/70  & 47/70 & 46/70 \\
\bottomrule
\end{tabular}
\end{table}

The frozen analysis resamples the 14 tasks for 20,000 bootstrap draws. Exact analyzer/verifier details, including the reconciled RNG seed, are in the reviewer artifact; the paper reports the pre-specified task-level procedure rather than treating 420 runs as independent observations.

\subsection{Continuation explains why a block is not the final outcome}
The continuation diagnostic asks what happens after a privileged proposal is blocked. Table~\ref{tab:app_continuation} keeps the two directions together because either one alone would give a misleading picture of intervention.

\begin{table}[!ht]
\centering
\caption{Continuation after blocked privileged proposals under CONFLICT. Blocking can recover the authorized effect or lose it, depending on what the agent does next.}
\label{tab:app_continuation}
\footnotesize
\begin{tabularx}{\columnwidth}{@{}P{0.29\columnwidth}C{0.10\columnwidth}Y@{}}
\toprule
Blocked proposal & Count & Final protected outcome \\
\midrule
Selected unauthorized & 13 & 12/13 later recover an authorization-equivalent effect (PAEF success) \\
Authorization-equivalent & 9 & 9/9 end with PAEF failure \\
\bottomrule
\end{tabularx}
\end{table}

The first row shows that blocking an unauthorized proposal need not determine the final effect; the second shows the opposite risk. The common lesson is the one used in the main text: a block is a state transition, not the final security outcome.

\subsection{A defense intervention can also change later inspection}
The source-and-trace audit in Figure~\ref{fig:live}c found an implementation-specific state transition in the tested AttriGuard configuration. A qualifying defense-generated all-empty tool-result batch sets the next-audit skip state; a following tool-call batch can then execute before the ordinary attenuation/shadow/judge audit. Mixed result batches do not satisfy the qualifying condition.

Table~\ref{tab:app_audit_ladder} keeps the four denominators separate because they answer different questions.

\begin{table}[!ht]
\centering
\caption{Denominator ladder for the later-inspection finding. The rows are different populations and should not be collapsed into a single bypass or exploit rate.}
\label{tab:app_audit_ladder}
\footnotesize
\begin{tabularx}{\columnwidth}{@{}Yr@{}}
\toprule
Population & Observed \\
\midrule
Defended runs entering the qualifying state & 44/210 \\
Defended runs later using that state & 22/210 \\
Defended privileged calls without ordinary audit & 18/168 \\
Unaudited privileged calls immediately after qualifying predecessor & 18/18 \\
\bottomrule
\end{tabularx}
\end{table}

The 18 unaudited privileged calls have mixed outcomes: 15 preserve an authorized or authorization-equivalent effect, two realize the selected unauthorized effect, and one produces neither. This is why the paper treats the mechanism as dual-use, source-bound evidence rather than an exploit-rate result.

A deterministic isolation provides the stronger mechanism check. Table~\ref{tab:app_isolation} shows the three conditions: the following privileged call is held fixed while the predecessor/state condition changes, and a local patch then restores ordinary inspection.

\begin{table}[!ht]
\centering
\caption{Deterministic isolation of the later-audit transition with the following privileged call held fixed.}
\label{tab:app_isolation}
\footnotesize
\begin{tabularx}{\columnwidth}{@{}P{0.35\columnwidth}C{0.20\columnwidth}Y@{}}
\toprule
Condition & Next audit & Route \\
\midrule
A: baseline predecessor & runs & ordinary inspected path \\
B: qualifying empty result & skipped & audit-skipped next batch \\
C': local patch, same next call & runs & inspection restored \\
\bottomrule
\end{tabularx}
\end{table}

This isolation supports the implementation mechanism but does not establish a general vulnerability or a universally correct upstream fix. The full source binding, trace records, and representative CASE\_001 runs remain in the artifact, where their different preceding trajectories can be inspected directly. Responsible disclosure has been completed.

\section{Why Task Success Can Hide a Changed Protected Action}
\label{app:replay}

Table~\ref{tab:evaluation} shows that downstream task success and immediate protected-action/effect fidelity can disagree. Replay isolates that distinction at one privileged decision. For each model--decision pair, the history is frozen immediately before the privileged action; only that next action is regenerated and executed. The original later tool calls and final answer are then replayed for the downstream utility check. The experiment therefore asks whether later continuation can hide an earlier action/effect change.

The 390 generations are five stability repeats over 78 model-by-decision pairs, not 390 independent cases. Table~\ref{tab:app_replay_cross} gives the complete majority-level cross-tab by model. The key disagreement is the lower-left column: 23/78 cells pass downstream while failing immediate action/effect fidelity. In 22/23 of those cells, the model still calls the intended tool.

\begin{table}[!ht]
\centering
\caption{Replay majority cross-tabs behind Table~\ref{tab:evaluation}. The third row is the evaluation gap: downstream success despite an immediate action/effect failure.}
\label{tab:app_replay_cross}
\footnotesize
\setlength{\tabcolsep}{3.2pt}
\begin{tabular}{@{}lrrrr@{}}
\toprule
Majority outcome & Llama & Gemma & Qwen & Total \\
\midrule
Immediate pass / downstream pass & 10 & 7 & 11 & 28 \\
Immediate pass / downstream fail & 0 & 0 & 0 & 0 \\
Immediate fail / downstream pass & 7 & 8 & 8 & 23 \\
Immediate fail / downstream fail & 9 & 11 & 7 & 27 \\
\bottomrule
\end{tabular}
\end{table}

The corrected action-local oracle compares the reconstructed post-action environment and canonicalizes only benchmark-generated timestamp values needed for stable equality. Superseded replay/oracle branches remain in the artifact as provenance; they are not part of the reviewer-facing scientific result.

\section{Reproducibility and Claim-to-Artifact Map}
\label{app:repro}

The Open Science appendix gives the anonymous artifact access point. This appendix provides navigation rather than another results section: it tells a reviewer which frozen evidence and deterministic check correspond to each paper-bearing claim. Provider transcripts, private paths, full row-level ledgers, large integrity manifests, and superseded engineering history stay in the artifact.

\paragraph{Natural relevance (Table~\ref{tab:natural}).}
The reviewer package contains the corrected natural result and decision ledger. A deterministic aggregation script recomputes the specified-versus-delegated comparison and its task-level uncertainty.

\paragraph{Source relocation and construct qualification (Figures~\ref{fig:relocation}--\ref{fig:construct}).}
The 24 frozen base instances, scorer condition scores, and matched N3 projection reproduce the USER--TOOL contrasts, support shifts, endpoint comparison, displacement comparison, and confidence intervals.

\paragraph{Threshold policy (Figure~\ref{fig:threshold} and Table~\ref{tab:threshold}).}
The artifact stores the complete frontier CSVs rather than a manuscript-sized sample. Verification regenerates every breakpoint/interval row and checks the reported anchor operating points.

\paragraph{AgentWatcher and AttriGuard (Figure~\ref{fig:architecture}).}
The paired AgentWatcher gate outputs reproduce the aligned/conflict 2$\times$2. The AttriGuard repeated-result ledger and route analysis reproduce the aggregate block endpoint before the exact-survival/later-review decomposition.

\paragraph{Live execution (Figure~\ref{fig:live}).}
Frozen run rows plus the pre-specified analyzer reproduce selected-ALT realization, PAEF, the task-level interaction, and continuation outcomes without re-running the agent.

\paragraph{Later inspection (Figure~\ref{fig:live}c).}
The source snapshot, live traces, and deterministic isolation reproduce the four-denominator ladder and the source-bound state transition. Raw trace bundles remain artifact-only because the paper needs the mechanism and its boundary, not full transcripts.

\paragraph{Evaluation fidelity (Table~\ref{tab:evaluation}).}
The replay joint result and per-model outputs reproduce the immediate-versus-downstream cross-tabs and the 23/78 majority-level disagreement.

Model-dependent full reruns may require third-party credentials or licensed components. The reviewer artifact therefore preserves frozen inputs and outputs so every reported statistic can be audited without reviewer credentials. The package must remove author usernames, institution/server names, private filesystem paths, credentials, tracking links, and identity-bearing Git metadata before submission.

\end{document}